\documentclass[showpacs,amsmath,amssymb,twocolumn,prl,superscriptaddress,notitlepage]{revtex4-1}

\usepackage{qcircuit} 
\usepackage[dvips]{graphicx}
\usepackage{amsmath,amssymb,amsthm,mathrsfs,amsfonts,dsfont}
\usepackage{subfigure, epsfig}
\usepackage{braket}
\usepackage{bm}
\usepackage{enumerate}
\usepackage{color}
\usepackage{graphicx}
\usepackage{algorithm}
\usepackage[noend]{algpseudocode}

\newcommand{\red}[1]{\textcolor{black}{#1}}

\newcommand{\black}[1]{\textcolor{black}{#1}}

\newcommand{\tr}{\mathrm{Tr}}

\begin{document}

% --------------------  TITLE  --------------------

\title{Variational quantum simulation of general processes}

% ------------  AUTHORS AND AFFILIATIONS ----------

\author{Suguru Endo }
\email{suguru.endo@materials.ox.ac.uk}
\affiliation{Department of Materials, University of Oxford, Parks Road, Oxford OX1 3PH, United Kingdom}

\author{Jinzhao Sun}
\affiliation{Clarendon Laboratory, University of Oxford, Parks Road, Oxford OX1 3PU, United Kingdom}

\author{Ying Li}
\affiliation{Graduate School of China Academy of Engineering Physics, Beijing 100193, China}

\author{Simon C. Benjamin}
\affiliation{Department of Materials, University of Oxford, Parks Road, Oxford OX1 3PH, United Kingdom}

\author{Xiao Yuan}
\email{xiao.yuan.ph@gmail.com}
\affiliation{Department of Materials, University of Oxford, Parks Road, Oxford OX1 3PH, United Kingdom}

% --------------------  ABSTRACT  --------------------

\begin{abstract}
{Variational quantum algorithms have been proposed to solve static and dynamic problems of closed many-body quantum systems.
Here we investigate variational quantum simulation of three general types of tasks---generalised time evolution with a non-Hermitian Hamiltonian, linear algebra problems, and open quantum system dynamics. 
The algorithm for generalised time evolution provides a unified framework for variational quantum simulation. 
In particular, we show its application  in solving linear systems of equations and matrix-vector multiplications by converting these algebraic problems into generalised time evolution. Meanwhile, assuming a tensor product structure of the matrices, we also propose another variational approach for these two tasks by combining variational real and imaginary time evolution. 
Finally, we introduce variational quantum simulation for open system dynamics. We variationally implement the stochastic Schr\"odinger equation, which consists of dissipative evolution and stochastic jump processes. 
}
{We numerically test the algorithm with a six-qubit 2D transverse field Ising model under dissipation. }
\end{abstract}

\maketitle

\emph{Introduction.---}The variational method is a powerful classical tool for simulating many-body quantum systems~\cite{BALIAN198829,RevModPhys.71.463,PhysRevLett.107.070601,SHI2018245,vanderstraeten2019tangent}. The core idea is based on the intuition that physical states with low energy belong to a  small manifold of the whole Hilbert space. As quantum circuits can efficiently prepare states that may not be efficiently represented classically, the variational method has been recently generalised to the quantum regime with trial states efficiently prepared by a quantum circuit and information extracted from a coherent measurement of the state~\cite{farhi2014quantum,peruzzo2014variational,wang2015quantum,PRXH2,PhysRevA.95.020501,VQETheoryNJP,PhysRevLett.118.100503,Li2017,PhysRevX.8.011021,Santagatieaap9646,kandala2017hardware,kandala2018extending,PhysRevX.8.031022,romero2018strategies,mcardle2019variational,jones2019variational,higgott2019variational,santagati2018witnessing,mcclean2017hybrid,colless2018computation,kokail2019self}.
The trial state in variational quantum algorithms can be prepared with shallow quantum circuits~\cite{kassal2011simulating,C2CP23700H,whaley2014quantum,mcardle2020quantum}, which is robust to a certain amount of device noise and is compatible with near-term Noisy Intermediate Scale Quantum (NISQ) hardware~\cite{preskill2018quantum}.
Variational quantum algorithms can be utilised for efficiently finding energy spectra~\cite{peruzzo2014variational,romero2018strategies,wang2015quantum,PRXH2,PhysRevA.95.020501,VQETheoryNJP,PhysRevLett.118.100503, jones2019variational,higgott2019variational,santagati2018witnessing,mcclean2017hybrid,colless2018computation,nakanishi2019subspace} and simulating real time Schr\"odinger evolution~\cite{Li2017,heya2019subspace} of closed systems. 
Although quantum circuits are unitary operations, the variational algorithm is not limited to energy minimisation and unitary processes and it can be used to simulate dissipative imaginary time evolution that cannot be \red{straightforwardly} mapped to unitary gates~\cite{mcardle2019variational,yuan2019theory}.

In this work, we study the capability of variational quantum algorithms and show that they are not limited to these applications. \red{First, we introduce  a variational quantum algorithm for simulating the generalised time evolution defined in Eq.~\eqref{Eq:general1} below.
Our algorithm can be regarded as a unified framework, which incorporates the special cases of  real and imaginary time evolutions~\cite{Li2017,mcardle2019variational,yuan2019theory}, non-Hermitian quantum mechanics~\cite{PhysRevLett.77.570,moiseyev_2011,bagarello2015non} that describes nonequilibrium processes~\cite{PhysRevLett.75.1883}, parity-time symmetric Hamiltonians~\cite{PhysRevLett.80.5243,bender2007making, bender2007faster}, open quantum systems~\cite{rotter2009non}, general first order differential equations, etc. }

\black{Next we apply the variational method for solving linear algebra problems, such as linear systems of equations and matrix-vector multiplications, important tasks in machine learning and optimisation~\cite{rasmussen2004gaussian, nasrabadi2007pattern}.} Many algorithms have been developed for linear systems of equations with universal quantum computers 
\cite{PhysRevLett.103.150502,doi:10.1137/16M1087072, ambainis2012variable, PhysRevLett.110.250504, PhysRevLett.120.050502, chakraborty_et_al:LIPIcs:2019:10609, gilyen2019quantum, PhysRevLett.122.060504}, which have profound applications in quantum machine learning~\cite{PhysRevLett.113.130503,lloyd2014quantum,biamonte2017quantum,2018arXiv180301601S,mitarai2018quantum}.  However they generally require deep circuits that rely on fault tolerant quantum computers.
\black{In this work, we introduce two types of variational quantum algorithms for the two linear algebra problems. For the first type, we consider general sparse matrices and show how solutions of the problems can be converted into generalised time evolutions, which can be variationally simulated. For the second type, we consider special matrices that are products of small matrices acting on a constant number of qubits, and use variational real and imaginary time evolution to find solutions.}

Finally, we combine the developed variational algorithms to simulate the evolution of open quantum systems~\cite{lindblad1976generators,breuer2002theory,gardiner2000quantum}. \black{Simulating the evolution of general open quantum systems is of great importance for understanding any quantum system that interacts with an environment. Existing quantum algorithms~\cite{wang2011quantum,bacon2001universal,sweke2015universal,sweke2016digital,cleve2016efficient,childs2017efficient} for simulating open quantum systems generally require deep quantum circuits. In this work, we consider the description of open system dynamics via the stochastic Schr\"odinger  equation, whose evolution can be regarded as an average of wave functions that undergo a continuous measurement induced from the environment~\cite{carmichael1993quantum,gardiner2000quantum}.}
The evolution of each wave function is composed of two processes that can be both simulated with variational algorithms: It may continuously evolve under the generalised time evolution with the system Hamiltonian and the damping effect due to continuous measurement; alternatively the state discontinuously jumps according to the measurement results. The continuous process can be described by the generalised time evolution, and the jump process is a matrix-vector multiplication process. 
Therefore our algorithm is compatible with shallow circuits and NISQ hardware. \\

\emph{Generalised time evolution.---}We first consider variational quantum simulation of generalised time evolution,
\begin{equation} \label{Eq:general1}
\black{B(t)\frac{d}{dt} \ket{v(t)} =  \ket{dv(t)}}.
\end{equation}
Here  $\ket{dv(t)}=\sum_j A_j(t) \ket{v'_j(t)}$, $A_j(t)$ \black{and $B(t)$} are general time dependent sparse (non-Hermitian) operators, $\ket{v(t)}$ is the system state, and each of $\ket{v'_j(t)}$ can be either  $\ket{v(t)}$ or any known state that can be efficiently prepared with a quantum circuit. 
The states $\ket{v(t)}$ and $\ket{v_j'(t)}$ can be (un)normalised states  $\ket{v(t)}=\alpha(t) \ket{\psi(t)}$, $\ket{v'_j(t)}=\alpha'_j(t) \ket{\psi'_j(t)}$ with normalisation factors $\alpha(t)$ and $\alpha_j'(t)$, respectively. 
 In practice, we assume that $A_j(t)$ \black{($B(t)$)} can be decomposed as a linear combination of Pauli operators $A_j(t)=\sum_i \lambda_i^j(t) \sigma_i$ with complex coefficients $\lambda_i$ and a polynomial (with respect to the system size) number of tensor products of Pauli matrices \textcolor{black}{$\sigma_i=\bigotimes_{i_k} \sigma_{i_k}$ with $i_k$ denoting the $i_k$th qubit.}

In variational quantum simulation, instead of directly simulating the dynamics, we assume that the state can be represented by parameterised quantum states $\ket{v(\vec{\theta}(t))}=\alpha({\vec{\theta}_0(t)}) \ket{\varphi(\vec{\theta}_1(t))}$ with $\vec{\theta}:=(\vec{\theta}_0, \vec{\theta}_1)$. Then we project the original evolution to the evolution of the parameters via McLachlan's principle~\cite{McLachlan}, 
\begin{equation}
\black{\min \bigg\|B(t)\frac{d}{dt}\ket{v(\vec{\theta}(t))} - \sum_j A_j(t) \ket{v_j'(t)}\bigg\|,}
\end{equation}
where $\|\ket{\psi}\| = \sqrt{\braket{\psi|\psi}}$ and the minimisation is over the derivative of the parameters.
By minimising the distance between the true evolution and the evolution of the parameterised state, we find the equation of parameters as
\begin{equation}\label{Eq:MVCEvol}
\sum_j \tilde{M}_{k,j}\dot{\theta}_j=\tilde{V}_k,
\end{equation}
where $\dot{\theta}_j=d\theta_j/dt$ and the coefficients are linear sums of state overlaps that can be efficiently measured with quantum circuits~\cite{mitarai2019methodology}. {We specify the detailed derivation, expression of the coefficients, the quantum circuits\black{, and a detailed resource estimation of the algorithm} in Supplementary Materials~\cite{NoteX}.} 
{Here we consider two examples with $B(t)=1$ and $\ket{dv(t)} = -iH\ket{v(t)}$ or $\ket{dv(t)} = -(H-\bra{v(t)}{H}\ket{v(t)})\ket{v(t)}$, corresponding to real and imaginary time evolution~\cite{Li2017,mcardle2019variational,yuan2019theory}, respectively. Compared to previous works studying real and imaginary time dynamics~\cite{Li2017,mcardle2019variational,yuan2019theory}, our algorithm considers a much more general setting of first-order differential equations.}
{Therefore, it not only unifies previous results in the general setting, but provides the basis for solving general problems as we discuss below.}
\\

\emph{Variational algorithms for linear algebra.---}\black{Now consider linear algebra problems of solving linear systems of equations and matrix-vector multiplications.}
For a sparse square matrix $\mathcal{M}$ and a state vector $\ket{v_0}$, we aim to find
\begin{equation}
\begin{aligned}
	\ket{v_{\mathcal{M}^{-1}}}&={\mathcal{M}^{-1} \ket{v_0}}\quad{\mathrm{or}} \quad \ket{v_{\mathcal M}}={\mathcal{M} \ket{v_0}}.
\end{aligned}
\end{equation}
\black{We introduce two types of algorithms where the first type is more general and the second type is more efficient with assumptions of the matrix. Here we take matrix multiplication as an example and the derivation works similarly for linear equations, which can also be found in  Supplementary Materials \cite{NoteX}. }

For the first type, {the algorithm is based on converting the static algebraic problem into a dynamical process, evolving the initial vector $\ket{v_0}$ to the target state $\ket{v_{\mathcal M}}$. }
A natural evolution path is via a linear extrapolation between $\ket{v_0}$ and $\ket{v_{\mathcal M}}$ as $\ket{v(t)}=E(t)\ket{v_0}$ with  $E(t)={t}/{T} \cdot \mathcal{M}+\big(1-{t}/{T}\big)I$, $\ket{v(0)}=\ket{v_0}$, and $\ket{v(T)}=\ket{v_{\mathcal M}}$. Different evolution paths can be also considered. For example, in the conventional Hamiltonian simulation scenario, we have $\mathcal{M}=e^{-iHT}$ and it corresponds to an exponential extrapolation. \red{We also consider linear extrapolation between normalised states in Supplementary Materials~\cite{NoteX} and we leave the discussion of other possible evolution paths to future works}. Given the evolution via linear extrapolation, the time derivative equation of $\ket{v(t)}$ is
$
\frac{d}{d t}\ket{v(t)}=G\ket{v(0)}
$,
with $G=\big(\mathcal{M}-I\big)/T$. This corresponds to the case with $A(t)=G$, \black{$B(t)=1$}, and $\ket{v'(t)}=\ket{v(0)}$ in Eq.~\eqref{Eq:general1}, which can be variationally simulated.

\black{For the second type, we assume that $\mathcal{M}$ is given as a tensor product of matrices, $\mathcal{M}=\mathcal{M}_1\otimes \dots \otimes \mathcal{M}_L$, with each $\mathcal{M}_i$ acting on a small constant number of qubits. We can thus sequentially act $\mathcal M_i$ and focus on the realisation of each term. 
For each $\mathcal M\equiv\mathcal M_i$}, 
we first consider a singular value decomposition as
$\mathcal{M}=U D V$, 
with unitary matrices $U$, $V$ and diagonal matrix $D$ with
non-negative entries. 
\black{Now we show how to realise the matrix multiplication of the three matrices. Given a spectral decomposition of $U=\sum_j e^{i \lambda_j}\ket{\lambda_j}\bra{\lambda_j}$ with $\lambda_j \in \mathbb{R}$, we can represent it as $U=\mathrm{exp}(-iH^{U} T^U)$ with $H^U=-\sum_j \lambda_j/T^U \ket{\lambda_j}\bra{\lambda_j}$ and $T^U>0$, which can be realised by evolving the state with Hamiltonian $H^{U}$ for time $T^U$ via variational real time simulation~\cite{Li2017}. The realisation is similar for $V$.
To realise the diagonal matrix $D=\sum_j a_j \ket{j} \bra{j}$, we approximate it as $D \approx \mathrm{exp} (-H^{D} T^D)$ with $-H^{D}T = \sum_{a_j\neq 0} \mathrm{log} (a_j) \ket{j} \bra{j} -\alpha  \sum_{a_j=0}  \ket{j}\bra{j}$ and a constant $\alpha = O\left(\log\left({1}/{\varepsilon_D}\right)+\log\textrm{Poly}(n)\right)$ that ensures an accuracy of $\varepsilon_D$ of the approximation with $n$ qubits.
We refer to Supplementary Materials for a detailed discussion~\cite{NoteX}.} Therefore, we can define an unnormalised imaginary time evolution 
%\begin{align}
$\frac{d \ket{v(\tau)}}{d\tau}= -H^{D} \ket{v(\tau)}$,
%\label{imaginary}
%\end{align}
so that the initial vector $\ket{v_0}$ at $\tau=0$  is evolved to $D\ket{v_0}$ at $\tau=T$. 
\black{Note that even though the second method assumes a tensor structure of $\mathcal M$, the matrix multiplication process can still be classically hard when the input state is a general multipartite entangled state. Such a case is practically relevant as we shortly discuss for its application in simulating jump processes of the stochastic Schr\"odinger  equation.} \\

\emph{Open system simulation.---}{Here we show quantum simulation of open system dynamics. Conventional algorithms for this task generally require a deep circuit~\cite{wang2011quantum,bacon2001universal,sweke2015universal,sweke2016digital,cleve2016efficient,childs2017efficient}. The recently proposed variational approach~\cite{yuan2019theory} also needs two copies of the purified quantum state, thus requiring a number of qubits that is four times that of the system size. 
Our method, based on variational algorithms for generalised time evolution and matrix multiplication and the stochastic Schr\"odinger equation, instead only requires to apply shallow circuits on a single copy of the state without purification. Our algorithm thus extensively alleviate the requirement of quantum simulation of open systems. }

Consider the Lindblad master equation,
\begin{align}\label{Eq:LindbladEquation}
\frac{d}{dt} \rho =-i[H, \rho]+\mathcal{L}\rho. 
\end{align}
where $H$ describes the system Hamiltonian  and $\mathcal{L}\rho =\sum_k \frac{1}{2}(2 L_k \rho L_k ^\dag -L_k^\dag L_k \rho -\rho L_k ^\dag L_k)$ describes the interaction with the environment
with Lindblad operators $L_k$. 
Instead of simulating the evolution of the density matrix, we consider its equivalent description by the stochastic Schr\"odinger equation, which averages trajectories of pure state evolution under continuous measurements~\cite{carmichael1993quantum,gardiner2000quantum}. 
Each single trajectory $\ket{\psi_c(t)}$ is described by
\begin{equation}\label{stocha1}
\begin{aligned}
	   d\ket{\psi_c(t)}=&\left(-iH-\frac{1}{2}\sum_{k}( L_k^\dag L_k -\braket{L_k^\dag L_k }  )\right) \ket{\psi_c(t)} dt  \\ &+\sum_{k }\left[\bigg( \frac{L_k \ket{\psi_c(t)}}{||L_k \ket{\psi_c(t)}||} - \ket{\psi_c(t)}\bigg)   dN_k\right],
\end{aligned}
\end{equation}
where $d\ket{\psi_c(t)}=\ket{\psi_c(t+dt)}-\ket{\psi_c(t)}$, $\braket{L_k^\dag L_k }=\bra{\psi_c(t)}L^\dag_k L_k \ket{\psi_c(t)}$, and $dN_k$ randomly takes either $0$ or $1$,  satisfing $dN_k dN_{k^\prime}=\delta_{k k^\prime} dN_k$ and $E[dN_k]=\bra{\psi_c(t)}L^\dag_k L_k \ket{\psi_c(t)}dt$. 
At each time $t$, we can assume a positive obervable valued measurement 
\textcolor{black}{$\{O_0=I- \sum_kL_k^\dag L_k  dt,\, O_k = L_k^\dag L_k dt \}$}  
happens. For measurement outcome $O_k$, the state discontinuously jumps to  $L_k \ket{\psi_c(t)}/ \|L_k \ket{\psi_c(t)} \|$ with probability $E[dN_k]$. And the total jump probability is $\gamma(t) = \sum_kE[dN_k]$.
For outcome $O_0$ with probability $1-\gamma(t)$, we have $dN_k=0 ~ \forall k$, and the state evolves under the generalised time evolution Eq.~\eqref{Eq:general1} with operator
\begin{equation}\label{Eq:Amaster}
	A = -iH-\frac{1}{2}\sum_{k}( L_k^\dag L_k -\braket{L_k^\dag L_k }  ),
\end{equation}
\textcolor{black}{and $B(t)=1$.}
Here, $-iH$ corresponds to the conventional real time Schr\"odinger evolution with Hamiltonian $H$ and the other terms can be understood as a normalised damping process.
Therefore, the whole process is composed of two parts: the continuous generalised time evolution and the quantum jump process.

The stochastic Schr\"odinger equation can be simulated with the Monto Carlo method. Suppose the state jumps at time $t$, then at time $t+\tau$, the probability $p(t+\tau)$ that the state does not jump is
$
	p(t+\tau) = e^{ -\Gamma(t, \tau)}
$,
with \textcolor{black}{$\Gamma(t, \tau)=\int_t ^{t+\tau} \gamma(t') dt^\prime $}. When a jump happens at time $t$, a uniform random number $q\in[0,1]$ is generated. Then the time of the next jump is determined by accumulating time $\tau$ until we have $p(t+\tau) =q$.
When a jump happens, a random number $q'\in[0,1]$ is generated to determine which jump operator to apply at each timestep. The state is updated to $L_k \ket{\psi_c(t)}/\|L_k \ket{\psi_c(t)} \|$ if $q'\in [\tilde {\gamma}_{k-1}(t), \tilde{\gamma}_k(t)]$, where 
\begin{align}\label{Eq:gammaJump}
\tilde{\gamma}_k(t)=\frac{\sum_{l=1}^k \bra{\psi_c(t)}L_l^\dag L_l\ket{\psi_c(t)}}{\sum_{l=1}^{N_L} \bra{\psi_c(t)}L_l^\dag L_l\ket{\psi_c(t)}},
\end{align}
and $N_L$ is the number of the Lindblad operators.
Considering discretised time with initial state $\ket{\psi_c(0)}$, the stochastic Schr\"odinger equation from time $0$ to $T$ can be simulated as follows.

\begin{algorithm}[H]
\begin{algorithmic}[1]
\State Set $\Gamma=0$ and generate a random number $q\in[0,1]$.
\For{$t=0:dt:T$} 
\If{$e^{-\Gamma} \textcolor{black}{\ge}  q$} 
\State Evolve the state $\ket{\psi_c(t)}$ under $A$ in Eq.~\eqref{Eq:Amaster}.
\State Calculate $\gamma(t)=\sum_k\bra{\psi_c(t)}L^\dag_k L_k \ket{\psi_c(t)}dt$. 
\State Update  $\Gamma = \Gamma + \gamma(t)$.
\Else 
\State Calculate $\tilde{\gamma}_k(t)$ in Eq.~\eqref{Eq:gammaJump}
\State Generate a random number $q'\in[0,1]$.
\If{$q'\in [\tilde {\gamma}_{k-1}(t), \tilde{\gamma}_k(t)]$} 
\State Update $\ket{\psi_c(t)}$ to $L_k \ket{\psi_c(t)}/\|L_k \ket{\psi_c(t)} \|$.
\EndIf
\State Reset $\Gamma=0$ and randomly generate $q\in[0,1]$.
\EndIf
\EndFor
\end{algorithmic}
	\caption{Stochastic evolution equation}\label{Al1}
\end{algorithm}

Now we show how to variationally simulate the stochastic Schr\"odinger equation, Algorithm~\ref{Al1}. 
Suppose the state $\ket{\psi_c(t)}$ at time $t$ can be represented by the parametrised state $\ket{\phi_c(\vec{\theta}(t))}$ prepared by a quantum computer. We can simulate step 4, i.e., the evolution under operator $A$ defined in Eq.~\eqref{Eq:Amaster}, with the algorithm for generalised time evolution. Specifically, we can evolve the parameters according to Eq.~\eqref{Eq:MVCEvol}
with
$\tilde{M}_{k,j}=\textrm{Re}\left(\frac{\partial \bra{\varphi(\vec{\theta}(t))}}{\partial \theta_k} \frac{\partial \ket{\varphi(\vec{\theta}(t))}}{\partial \theta_j}\right)$,
$\tilde{V}_k=\textrm{Re}\left(\bra{\varphi(\vec{\theta}(t))}(-iH-(L-\braket{L})) \frac{\partial \ket{\varphi(\vec{\theta}(t))}}{\partial \theta_k}\right)
$,
and $L=\frac{1}{2}\sum_k L_k^\dag L_k$. 
The values of $\gamma(t)$ and $\tilde{\gamma}_k(t)$ at step 5 and 8 are measured as expectation values of quantum states. The jump at step 11 is realised by the variational algorithms for matrix-vector multiplication. Especially, when considering $L_k$ as a product of operators of each qubits, it can be efficiently realised with the singular value decomposition method.    
In practice, we consider sparse Hamiltonian and Lindblad operators, therefore all the measurements can be efficiently evaluated. \black{We refer to Supplementary Materials for the discussion of the resource estimation~\cite{NoteX}.}\\

\begin{figure}[t]\centering
\includegraphics[width=0.55\columnwidth]{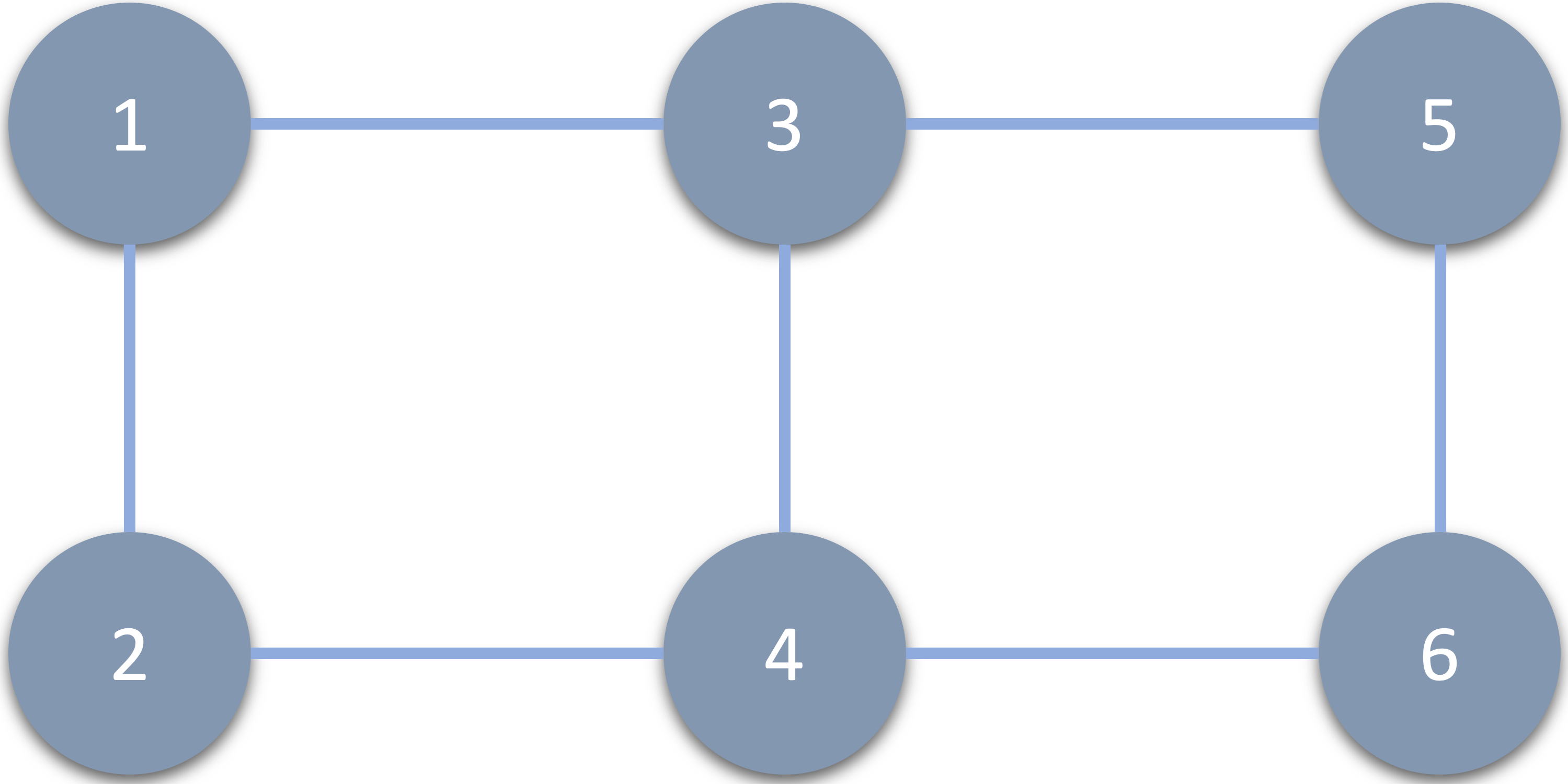}
\caption{Geometry of the 2D Ising model. We consider nearest-neighbor interactions for two connected qubits. }
\label{layout}
\end{figure}

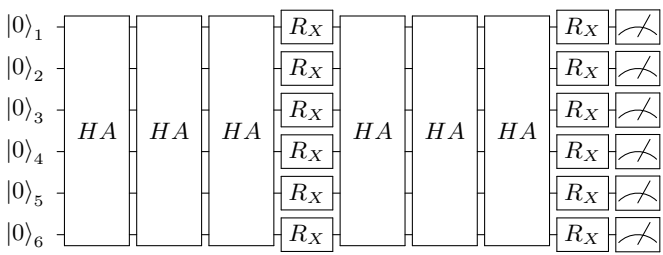
\begin{figure}[b]
\centering
\begin{align*}
\Qcircuit @C=.3em @R=.3em {
\lstick{\ket{0}_1}&\multigate{5}{HA}&\multigate{5}{HA}&\multigate{5}{HA}&\gate{R_X}&\multigate{5}{HA}&\multigate{5}{HA}&\multigate{5}{HA}&\gate{R_X}&\meter\\
\lstick{\ket{0}_2}&\ghost{HA}&\ghost{HA}&\ghost{HA}&\gate{R_X}&\ghost{HA}&\ghost{HA}&\ghost{HA}&\gate{R_X}&\meter\\
\lstick{\ket{0}_3}&\ghost{HA}&\ghost{HA}&\ghost{HA}&\gate{R_X}&\ghost{HA}&\ghost{HA}&\ghost{HA}&\gate{R_X}&\meter\\
\lstick{\ket{0}_4}&\ghost{HA}&\ghost{HA}&\ghost{HA}&\gate{R_X}&\ghost{HA}&\ghost{HA}&\ghost{HA}&\gate{R_X}&\meter\\
\lstick{\ket{0}_5}&\ghost{HA}&\ghost{HA}&\ghost{HA}&\gate{R_X}&\ghost{HA}&\ghost{HA}&\ghost{HA}&\gate{R_X}&\meter\\
\lstick{\ket{0}_6}&\ghost{HA}&\ghost{HA}&\ghost{HA}&\gate{R_X}&\ghost{HA}&\ghost{HA}&\ghost{HA}&\gate{R_X}&\meter\\
}
\end{align*}
\caption{
\textcolor{black}{The ansatz} consists of the Hamiltonian ansatz (HA) and single qubit rotations. Each HA is  
$
e^{\theta_1 (Z_5Z_6)}
e^{\theta_2 (Z_3Z_5+Z_4Z_6)}
e^{\theta_3 (Z_3Z_4)}
e^{\theta_4 (Z_1Z_3+Z_2Z_4)}
e^{\theta_5 (Z_1Z_2)}$ 
$
e^{\theta_6 (X_3+X_4)}
e^{\theta_7 (X_1+X_2+X_5+X_6)}$ and each single qubit gate $R_X$ is $e^{-i\theta X}$. With different parameters for different HA and single qubit gates, the ansatz has in total $54$ parameters. } \label{Fig:ansatz}
\end{figure}

\emph{Numerical simulation.---}We numerically test the variational algorithm  for simulating a {six-qubit 2D Ising model in a transverse field coupled to a Markovian bath~\cite{lee2012collective,ates2012dynamical,lesanovsky2013characterization,rose2016metastability}.
The Hamiltonian is
$H_I=J/4 \sum_{\langle ij \rangle}  Z_i Z_{j} +h_X\sum_{i=1}^6 X_i$, where Pauli operators $X_i$, $Y_i$, and $Z_i$ act on the $i^{\textrm{th}}$ spin and $\langle ij \rangle$ represents nearest neighbor pairs in Fig.~\ref{layout}.
The Lindblad term is $L_i = \sqrt{\gamma}\sigma^+_i$ with 
$\sigma_i^+= \ket{1}\bra{0}_i$ being the raising operator acting on the $i^{\textrm{th}}$ spin.
In our simulation, we set $J=1$, $h_X=1$,  and $\gamma=1$.  The initial state is prepared in a product state$\ket{\varphi(0)}=\ket{0}^{\otimes 6}$, and then the Hamiltonian $H_I$ is suddenly turned on, which drives the qubits out of equilibrium. We simulate both the ideal and dissipative evolution  from $t=0$ to $t=6$, and measure the normalised nearest-neighbor correlations $C=\sum_{\langle ij \rangle}Z_iZ_j/7$. 
We use the Hamiltonian ansatz (HA)~\cite{PhysRevA.92.042303} sandwiched with single qubit rotations as the trial state,  shown Fig.~\ref{Fig:ansatz}. The HA preserves the symmetry of the Hamiltonian and the single qubit rotations are introduced to break the symmetry, ensuring its capability for simulating the jump process.}

\begin{figure}[t]
\includegraphics[width=\columnwidth]{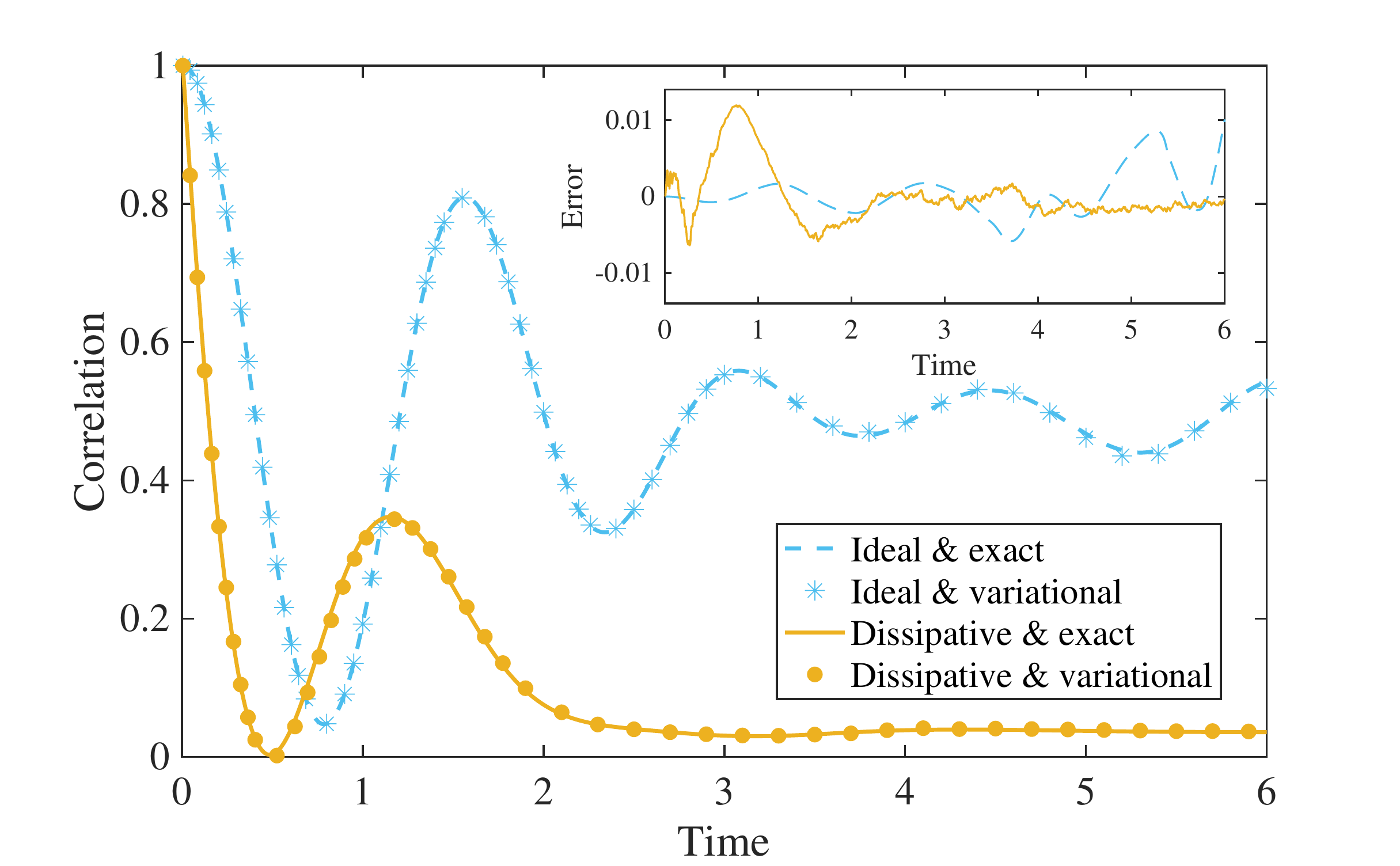}
\caption{Numerical simulation of ideal and dissipative evolution of the 2D Ising model.  We consider evolution from $t=0$ to $6$ with $\delta t =0.005$ and measure the nearest-neighbor correlations $C = \sum_{\langle ij \rangle}Z_iZ_j/7$. 
We repeat the stochastic process of Algorithm 1 with $N_{\mathrm{trial}}=2 \times 10^4$ times.  
The results from variational simulation agree well with the exact ones, with a maximal error  around 0.01 (see inset).
}
\label{kekka}
\end{figure}

We apply the algorithm in Ref.~\cite{Li2017} and our Algorithm~1 to simulate the ideal and dissipative evolution, respectively. 
We use the singular value decomposition method
to simulate quantum jumps in Algorithm~1. 
We decompose the jump operator $\sigma^+$ as $\ket{1}\bra{0}=\ket{1}\bra{1} X$ with $\ket{1}\bra{0} = UDV$,  $U=I$, $D=\ket{1}\bra{1}$, and $V=X$. 
To realise the $V$ operator, we set $H_V=X$ and $T_V=\pi/2$ such that $X=\exp(-iH_VT_V)$. Then we evolve the state under Hamiltonian $H_V$ for time $T_V$ with time step $\delta t_V = 0.01$ to have $X\ket{\varphi (\vec{\theta})}$. To realise $D=\ket{1}\bra{1}$, we set $H_D=\ket{0}\bra{0}$ and $T_D=6$ so that $D\approx\exp(-H_DT_D)$. Then we realise $D\ket{\varphi (\vec{\theta})}/\|D\ket{\varphi (\vec{\theta})}\|$ by using the normalised variational imaginary time evolution with total time $T_D$ and time step $\delta t_D= 0.1$.

{The simulation results are shown in Fig.~\ref{kekka}. We compare the dynamical nearest-neighbor correlation of the simulation result and the exact evolution. We can see that the simulation result agrees well with exact evolution for both ideal and dissipative cases, thus verifying the functioning of variational quantum algorithms. The simulation result indicates a dissipation induced phase transition, where similar phenomena has been experimentally investigated in Ref. \cite{raftery2014observation}. We expect our algorithm with medium-scale quantum hardware may be used for probing general interesting physics phenomena of many-body open systems.}  \\

\emph{Discussion.---}To summarise, we extend the variational quantum simulation method to general processes, including generalised time evolution, \black{and its application for solving linear algebra tasks and simulating the evolution of open quantum systems. }
Our algorithm for simulating the generalised time evolution can be applied to simulate non-Hermitian quantum mechanics~\cite{PhysRevLett.77.570,moiseyev_2011,bagarello2015non} including nonequilibrium processes~\cite{PhysRevLett.75.1883} and parity-time symmetric Hamiltonians~\cite{PhysRevLett.80.5243,bender2007making, bender2007faster}.
Especially, it is shown in Ref.~\cite{bender2007faster} that a quantum state can evolve to the target state faster with non-Hermitian parity-time symmetric Hamiltonians than the case with Hermitian Hamiltonians. Therefore, our variational algorithm for simulating the generalised time evolution  may also be useful for designing faster quantum computing algorithms.
\black{Recently, several other algorithms for linear algebra tasks have been proposed to be compatible with near-term quantum hardware~\cite{xu2019variational,bravo2019variational,an2019quantum,huang2019near}. A thorough comparison of these algorithms and ours could also be an interesting future work. }
{Our algorithm for simulating open systems only needs shallow circuits on a number of qubits matching the system size, thus it enables the possibility of investigating open system physics with near-term quantum computers.}
Finally the proposed algorithms are compatible with NISQ hardware and can be further combined with quantum error mitigation techniques~\cite{Li2017,samerrormitigation,bonet2018low,PhysRevLett.119.180509,endo2017practical,huo2018temporally,otten2018accounting,kandala2018extending,sun2020practical}.

\begin{acknowledgements}
\emph{Acknowledgements.---}This work is supported by the EPSRC National Quantum Technology Hub in Networked Quantum Information Technology (EP/M013243/1). SCB acknowledges support from the EU AQTION project. We acknowledge the use of the University of Oxford Advanced Research Computing (ARC) facility in carrying out this work.
SE is supported by Japan Student Services Organization (JASSO) Student Exchange Support Program (Graduate Scholarship for Degree Seeking
Students). 
YL is supported by National Natural Science Foundation of China (Grant No. 11875050) and NSAF (Grant No. U1930403).
JS acknowledges support from Chinese Scholarship Council. 
\end{acknowledgements}

\bibliographystyle{apsrev4-1}
\bibliography{bib}

\newpage
\widetext

\section*{Supplementary materials: Variational quantum simulation of general processes}

\section{Review: variational quantum simulation of real and imaginary time evolution}
\label{Sec:realimag}
We first review the variational quantum algorithms for simulating real and imaginary time evolution introduced in Ref.~\cite{Li2017} and~\cite{mcardle2019variational}, respectively, \textcolor{black}{as we generalise these algorithms to propose our new algorithm for general processes.} We refer the reader to Ref.~\cite{yuan2019theory} for a comprehensive derivation of quantum variational algorithms from various variational principles --- the Dirac and Frenkel variational principle, the McLachlan's variational principle, and the time-dependent variational principle.

The real time evolution is described by the Schr\"odinger equation, 
\begin{align}
\label{schrodinger}
\frac{d \ket{\psi(t)}}{dt}=-i H \ket{\psi(t)},
\end{align}
with Hermitian Hamiltonian $H$. Instead of directly simulating the real time dynamics with the Hamiltonian simulation algorithms~\cite{lloyd1996universal,suzuki1991general,childs2019faster,berry2015simulating,low2019hamiltonian}, the variational method assumes that the quantum state $\ket{\psi(t)}$ is prepared by a parametrised quantum circuit, $\ket{\varphi (\vec{\theta}(t))}=R_{N}(\theta_N)\dots R_{k}(\theta_k)\dots R_{1}(\theta_1)\ket{\bar{0}}$ with each gate $R_{k}(\theta_k)$ controlled by the real parameter $\theta_k$ and the reference state $\ket{\bar{0}}$. Here, we denote $\vec{\theta}= (\theta_1, \theta_2,\dots, \theta_N)$. 
According to McLachlan's variational principle~\cite{McLachlan}, the real time dynamics of $\ket{\psi(t)}$ can be mapped to the evolution of the parameters $\vec{\theta}(t)$ by minimising the distance between the ideal evolution and the evolution induced of the parametrised trial state, 
\begin{align}
\delta \|(\partial/\partial t + iH )\ket{\varphi (\vec{\theta}(t))}  \|=0, 
\end{align}
where $\|\ket{\varphi} \|=\sqrt{\braket{\varphi|\varphi}}$. The solution is
\begin{align}
\sum_j M_{k,j}\dot{\theta}_j=V_k, 
\end{align}
with coefficients
\begin{equation}\label{Eq:MV}
\begin{aligned}
	  M_{k,j}&=\mathrm{Re}\bigg(\frac{\partial \bra{\varphi(\vec{\theta}(t))}}{\partial \theta_k}\frac{\partial \ket{\varphi(\vec{\theta}(t))}}{\partial \theta_j}\bigg),\\
V_k&=\mathrm{Im}\bigg(\bra{\varphi(\vec{\theta}(t))}H \frac{\partial \ket{\varphi(\vec{\theta}(t))}}{\partial \theta_k}\bigg). 
\end{aligned}
\end{equation}

For imaginary time evolution, the normalised Wick-rotated Shr\"odinger equation is obtained by replacing $t=i\tau$ in Eq.~\eqref{schrodinger},
\begin{align}
\frac{d \ket{\psi(\tau)}}{d\tau}=- (H - \bra{\psi(\tau)}H \ket{\psi(\tau)})\ket{\psi(\tau)}.
\end{align}
Applying a similar procedure for real time evolution, the imaginary time evolution is mapped to the evolution of the parameters via McLachlan's principle,
\begin{align}
\delta \|(\partial/\partial \tau +H-\braket{H})\ket{\varphi(\vec{\theta}(t))} \|=0. 
\end{align}
The  evolution of the parameters is
\begin{align}
\sum_j M_{k,j} \dot{\theta}_j= C_k,
\end{align}
with $M$ given in Eq.~\eqref{Eq:MV} and $C$ defined by 
\begin{align}
C_k= -\mathrm{Re} \left(\bra{\varphi(\vec{\theta}(t))}H \frac{\partial \ket{\varphi(\vec{\theta}(t))}}{\partial \theta_k}\right).
\end{align}

The $M$, $V$, and $C$ terms can be efficiently measured with quantum circuits. Considering gate based circuits, the derivative of the each parameterised gate can be expressed as 
\begin{eqnarray}
\frac{\partial R_{k}}{\partial \theta_{k}} = \sum_{i} g_{k,i} R_{k} \sigma_{k,i}, 
\label{eq:expansion}
\end{eqnarray}
where $\sigma_{k,i}$ are unitary operators and $g_{k,i}$ are complex coefficients. The derivative of the trial state can be written as
\begin{eqnarray}\label{Eq:partialstate}
\frac{\partial \ket{\varphi(\vec{\theta}(t))}}{\partial \theta_{k}} = \sum_{i} g_{k,i} R_{k,i} \ket{\bar{0}},
\end{eqnarray}
where
\begin{eqnarray}
R_{k,i}= R_{N} R_{N-1} \cdots R_{k+1} R_{k} \sigma_{k,i} \cdots R_{2} R_{1}.
\end{eqnarray}
The $M_{k,j}$ terms can be expressed as
\begin{eqnarray}
M_{k,j} = \sum_{i,j}\Re \left(
g^*_{k,p}g_{j,q} \bra{\bar{0}} R^\dag_{k,p} R_{j,q} \ket{\bar{0}}
\right).
\label{eq:M}
\end{eqnarray}
Similarly, considering sparse Hamiltonian with decomposition $H=\sum_j \lambda_j \sigma_j$, $\lambda_j \ \in \mathbb{R}$,  we have $C_k$ and $V_k$ as 
\begin{equation}\label{eq:V}
\begin{aligned}
V_k &= \sum_{i,j} \mathrm{Re} \left(i
g^*_{k,i}\lambda_{j} \bra{\bar{0}} R^\dag_{k,i}\sigma_{j}R \ket{\bar{0}}
 \right), \\
C_{k}& = -\sum_{i,j} \mathrm{Re} \left(
g^*_{k,i}\lambda_{j} \bra{\bar{0}} R^\dag_{k,i}\sigma_{j}R \ket{\bar{0}}
 \right), 
\end{aligned}
\end{equation}
All the $M$, $C$, and $V$ terms can be written in the form
$$
a \mathrm{Re} \left( e^{i\theta} \bra{\bar{0}} U \ket{\bar{0}} \right),
$$
where $a, \theta \in \mathbb{R} $ depend on the coefficients, and $U$ is a unitary operator of either $R^\dag_{k,p}R_{j,q}$ or $R^\dag_{k,i}\sigma_{j}R$. We can calculate $M$, $C$, and $V$ by using the quantum circuit shown in Fig.~\ref{Fig:circuitPrac}. 

\begin{figure*}[h!t]
\begin{align*}
\Qcircuit @C=1em @R=.7em {
\lstick{(\ket{0}+e^{i\theta}\ket{1})/\sqrt{2}}&\qw&\qw&\gate{X}&\ctrl{2}&\gate{X}&\qw&\qw&\ctrl{2}&\gate{H}& \meter\\
&&...&&&&...&\\
\lstick{\ket{\bar{0}}}&\gate{R_1}&\qw&\gate{R_{k-1}}&\gate{\sigma_{k,p}}&\gate{R_{k}}&\qw&\gate{R_{j-1}}&\gate{\sigma_{j,q}}&\qw&\qw\\
}
\end{align*}
(a)
\begin{align*}
\Qcircuit @C=1em @R=.7em {
\lstick{(\ket{0}+e^{i\theta}\ket{1})/\sqrt{2}}&\qw&\qw&\gate{X}&\ctrl{2}&\gate{X}&\qw&\qw&\ctrl{2}&\gate{H}& \meter\\
&&...&&&&...&\\
\lstick{\ket{\bar{0}}}&\gate{R_1}&\qw&\gate{R_{k-1}}&\gate{\sigma_{k,i}}&\gate{R_{k}}&\qw&\gate{R_{N}}&\gate{\sigma_j}&\qw&\qw\\
}
\end{align*}
(b)
\caption{Quantum circuits that evaluate (a) $\mathrm{Re}(e^{i\theta}\bra{\bar{0}}R_{k,p}^\dag R_{j,q}\ket{\bar{0}})$ and (b) $\mathrm{Re}(e^{i\theta}\bra{\bar{0}} R_{k,i}^\dag \sigma_j R\ket{\bar{0}})$. 
}\label{Fig:circuitPrac}
\end{figure*}

\section{Derivation for variational simulation of generalised time evolution equation}
Now, we consider variational simulation of the  generalised time evolution equation,
\begin{equation}
B(t)\frac{d}{dt}\ket{v(t)}=\sum_j A_j(t) \ket{v'_j(t)}.
\label{gen}
\end{equation}

By parametrising $\ket{v(t)}$ and $ \ket{v'_j(t)}$ as $\ket{v(\vec{\theta}(t))}$ and $\ket{v'(\vec{\theta}'_j(t))}$, with McLachlan's principle, we have 
\begin{equation}
\delta \left \| B(t)\sum_i \frac{\partial  \ket{v(\vec{\theta}(t))}}{ \partial \theta_i} \dot{\theta}_i -\sum_j A_j(t) \ket{v'(\vec{\theta}'_j(t))} \right\|	=0.
\end{equation}
This is equivalent to 
\begin{equation}
\begin{aligned}
&\frac{\partial}{\partial \dot{\theta}_k} \left \| B(t)\sum_i \frac{\partial  \ket{v(\vec{\theta}(t))}}{ \partial \theta_i} \dot{\theta}_i -\sum_j A_j(t) \ket{v'(\vec{\theta}'_j(t))} \right\| \\ 
&= \frac{\partial}{\partial \dot{\theta}_k} \left( \sum_i \frac{\partial  \bra{v(\vec{\theta}(t))}}{ \partial \theta_i} \dot{\theta}_i B^\dag(t)-\sum_j  \bra{v'(\vec{\theta}'_j(t))} A_j^\dag(t) \right) \\
&\left(\sum_l B(t)\frac{\partial  \ket{v(\vec{\theta}(t))}}{ \partial \theta_l} \dot{\theta}_l -\sum_j A_j(t) \ket{v'(\vec{\theta}'_j(t))} \right)=0 ~ \forall k.
\end{aligned}	
\end{equation}
Hence, we have 
\begin{equation}
\begin{aligned}
&\sum_j \left(\frac{\partial \bra{v(\vec{\theta}(t))}}{\partial \theta_k}B^\dag(t)B(t)\frac{\partial \ket{v(\vec{\theta}(t))}}{\partial \theta_j} +\frac{\partial \bra{v(\vec{\theta}(t))}}{\partial \theta_j}B^\dag(t)B(t)\frac{\partial \ket{v(\vec{\theta}(t))}}{\partial \theta_k} 
\right) \dot{\theta}_j \\
&=\sum_j \frac{\partial \bra{v(\vec{\theta}(t))}}{\partial \theta_k} B^\dag(t)A_j(t) \ket{v'_j (\vec{\theta}_j'(t))}+h.c,
\end{aligned}
\end{equation}
which leads to Eq.~(\ref{Eq:MVCEvol}) in the main text.

\begin{align}
\sum_j \tilde{M}_{k,j} \dot{\theta}_j = \tilde{V}_k.	 \nonumber
\end{align}
By substituting $\ket{v(\vec{\theta}(t))}=\alpha(\vec{\theta_0}(t)) \ket{\varphi(\vec{\theta}_1(t))}$ and $\ket{v'(\vec{\theta}'_j(t))}=\alpha'(\vec{\theta}_{0j}' (t)) \ket{\varphi(\vec{\theta}_{1j}'(t))}$, we have
\begin{equation}\label{Eq:MVCmain}
\begin{aligned}
	 & \tilde{M}_{k,j}=\mathrm{Re}\left(|\alpha({\vec{\theta}_0(t)})|^2 \frac{\partial \bra{\varphi(\vec{\theta_1}(t))}}{\partial \theta_k} B^\dag(t)B(t)\frac{\partial \ket{\varphi(\vec{\theta_1}(t))}}{\partial \theta_j}\right)\\
&+\mathrm{Re} \left(\frac{\partial \alpha^* (\vec{\theta}_0 (t))}{\partial \theta_k} \alpha (\vec{\theta}_0(t)) \bra{\varphi (\vec{\theta}_1(t))}B(t)\frac{\partial \ket{\varphi (\vec{\theta}_1(t))}}{\partial \theta_j}  \right) \\
&+\mathrm{Re} \left(\frac{\partial \alpha^* (\vec{\theta}_0 (t))}{\partial \theta_j} \alpha (\vec{\theta}_0(t)) \bra{\varphi (\vec{\theta}_1(t))}B(t)\frac{\partial \ket{\varphi (\vec{\theta}_1(t))}}{\partial \theta_k}  \right)  \\
&+ \mathrm{Re}\left(\frac{\partial \alpha({\vec{\theta}_0(t)})}{\partial \theta_k} \frac{\partial \alpha^*({\vec{\theta}_0(t)})}{\partial \theta_j} \bra{\varphi(\vec{\theta}_1(t))} B^\dag(t)B(t) \ket{\varphi(\vec{\theta}_1(t))}
\right), \\ 
&\tilde{V}_k = \sum_j \mathrm{Re}\left(\frac{\partial \alpha^* (\vec{\theta}_0(t))}{\partial \theta_k} \alpha' (\vec{\theta}_{0 j}' (t))\bra{\varphi (\vec{\theta}_1(t))} B^\dag(t)
A_j(t) \ket{\varphi'_j (\vec{\theta}_{1j}'(t))}  \right). \\
&+\sum_j \mathrm{Re}\left(\alpha ^*({\vec{\theta}_0(t)}) \alpha'({\vec{\theta}_{0 j}'(t)}) \frac{\partial \bra{\varphi(\vec{\theta}_1(t))}}{\partial \theta_k} B^\dag(t)A_j(t) \ket{\varphi'_j(\vec{\theta}_{1j}'(t))} \right).
\end{aligned}
\end{equation}

The first term of $\tilde{M}_{k,j}$ can written as

\begin{align}
\sum_{i,q,l} \mathrm{Re}\bigg( |\alpha({\vec{\theta}_0(t)})|^2  g^*_{k,i}g_{j,q} \beta_l \bra{\bar{0}}R_{k,i}^\dag \sigma_l R_{j,q} \ket{\bar{0}}\bigg),	
\end{align}
where we set $B^\dag(t)B(t)=\sum_l \beta_l \sigma_l$, and $\sigma_l$ is a Pauli operator. Each term of $\tilde{M}_{k,j}$ can be written in the form of $a~\mathrm{Re}(e^{i \theta} \bra{\bar{0}}R_{k,i}^\dag \sigma_l R_{j,q} \ket{\bar{0}} )$, where $a, \theta \in \mathbb{R}$. The quantum circuit to compute this value is shown in Fig. \ref{tilmij}. It is worth mentioning that this quantum circuit only necessitates three controlled operations from an ancilla qubit. The second and third terms can be written as $\sum_{i j} \mathrm{Re}\big(\frac{\partial \alpha^* (\vec{\theta}_0 (t))}{\partial \theta_k} \alpha (\vec{\theta}_0(t)) \gamma_j g^*_{k,i}\bra{\bar{0}}R^\dag _{k,i} \sigma_j R \ket{\bar{0}}  \big)$, where we set $B(t)=\sum_j \gamma_j \sigma_j $. We can express each term in the form of $a~\mathrm{Re}(e^{i \theta} \bra{\bar{0}}R_{k,i}^\dag \sigma_l R \ket{\bar{0}} )$, which can be evaluated by using the quantum circuit shown in Fig. \ref{Fig:circuitPrac} (b). The fourth term can be simply computed by measuring the expectation value of $B^\dag(t) B(t)=\sum_l \beta_l \sigma_l$ for $\ket{\varphi(\vec{\theta}_1)}$. 

The term $\tilde{V}$ can be computed as follows. We denote $\tilde{U}_t$ and $\tilde{U}_t'$ to be the unitary circuit to prepare $\ket{\varphi (\vec{\theta}_1(t))}=\tilde{U}_t \ket{\bar{0}}$ and $\ket{\varphi'_j (\vec{\theta}_{1j}'(t))}=\tilde{U}_t' \ket{\bar{0}}$. Replacing $B^\dag(t)A_j(t)=\sum_l \Lambda^j_l(t) \sigma_l$, the first term of each $\tilde{V}_k$ can be written as $\sum_{jl} \mathrm{Re}\big( \alpha^*(\vec{\theta}_0(t)) \alpha'(\vec{\theta}'_{0j}(t)) \Lambda_l^j \bra{\bar{0}} \tilde{U}^\dag_t \sigma_l \tilde{U}'_t \ket{\bar{0}}  \big) $.  Each term can be expressed in the form of $a \mathrm{Re}\big(e^{i\theta} \bra{\bar{0}}\tilde{U}_t^\dag \sigma_l \tilde{U}'_t\ket{\bar{0}} \big)$ and computed using the quantum circuit shown in Fig.~\ref{Fig:qcirc}. Meanwhile, the second term can be expressed as $\sum_{ijl} \mathrm{Re}( \alpha^* (\vec{\theta}_0(t)) \alpha'(\vec{\theta}_{0j}(t)) \Lambda_l^j(t) g^*_{k,i}  \bra{\bar{0}}R^\dag_{k,i}\sigma_{l}\tilde{U}_t' \ket{\bar{0}})$, and each term can be written in the form of $a \mathrm{Re}\big(e^{i\theta} \bra{\bar{0}}R_{k,i}^\dag \sigma_l \tilde{U}_t' \ket{\bar{0}} \big)$, which can be computed by using the quantum circuit shown in Fig. \ref{Fig:qcirc2}. 

Note that, although the quantum circuits to evaluate $\tilde{V}$ generally need controlled $U$ operations, in the case $\ket{v'(\vec{\theta}'_{j}(t))}=\ket{v(\vec{\theta}(t))}$, e.g., open quantum simulation and solving linear equations, it can be computed by the quantum circuit shown in Fig. \ref{Fig:circuitPrac} (b), which only needs two controlled operations.

\begin{figure*}[h!t]
\begin{align*}
\Qcircuit @C=1em @R=.7em {
\lstick{(\ket{0}+e^{i\theta}\ket{1})/\sqrt{2}}&\qw&\qw&\qw&\ctrl{2}&\gate{X}&\qw&\qw&\ctrl{2}&\qw&\qw&\qw&\qw&\ctrl{2}&\gate{X}&\gate{H}&\meter\\
&&...&&&&...&&&&&... \\
\lstick{\ket{\bar{0}}}&\gate{R_1}&\qw&\gate{R_{k-1}}&\gate{\sigma_{k,i}}&\gate{R_{k}}&\qw&\gate{R_{j-1}}&\gate{\sigma_{j,q}}&\qw&\gate{R_{j+1}}&\qw&\gate{R_N}&\gate{\sigma_l}&\qw&\qw \\
}
\end{align*}

\caption{Quantum circuits that evaluate $\mathrm{Re}\bigg(g^*_{k,i}g_{j,q}\bra{\bar{0}}R_{k,i}^\dag \sigma_l R_{j,q} \ket{\bar{0}}\bigg)$.
}
\label{tilmij}
\end{figure*}

\begin{figure}[h!t]
\centering 
\begin{align*}
\Qcircuit 
 @C=1.0em @R=1.5em {
 \lstick{\ket{0}+e^{i \phi_l} \ket{1}} &\ctrl{1} & \ctrl{1} & \gate{X} & \ctrl{1} & \gate{X} & \gate{H}  & \meter   \\ 
\lstick{\ket{\bar{0}}} & \gate{\tilde{U}_t} & \gate{\sigma_l} & \qw & \gate{\tilde{U}_t'} & \qw &  \qw & \qw  }
\end{align*}
\caption{The quantum circuit for evaluating $\tilde{V}_k$.}
\label{Fig:qcirc}
\end{figure}

\begin{figure}[h!t]
\centering 
\begin{align*}
\Qcircuit 
 @C=1.0em @R=1.5em {
 \lstick{\ket{0}+e^{i \phi_l} \ket{1}} &\ctrl{1} & \ctrl{1} & \gate{X} & \ctrl{1} & \gate{X} & \gate{H}  & \meter   \\ 
\lstick{\ket{\bar{0}}} & \gate{R_{k,i}} & \gate{\sigma_l} & \qw & \gate{\tilde{U}_t} & \qw &  \qw & \qw  }
\end{align*}
\caption{The quantum circuit for evaluating $\tilde{V}_k$.}
\label{Fig:qcirc2}
\end{figure}

\section{Variational simulation for linear algebra tasks}
\subsection{Matrix evolution with normalised states}
Here, we also consider the case where we are only interested in the normalised final state {$\ket{\psi(t)}=\mathcal{M}\ket{v_0}/\|\mathcal{M}\ket{v_0}\|$. By extrapolating from $\ket{v_0}/\|\ket{v_0}\|$} to $\ket{\psi(t)}$, we can similarly have an evolution of the state $\ket{\psi(t)}$ as
\begin{equation}
	\ket{\psi(t)}=E'(t)\ket{\psi_0},
\end{equation}
with 
\begin{align}
E^\prime(t)= N(t) \bigg(\frac{t}{T} \mathcal{M} + (1-\frac{t}{T})  I \bigg),
\end{align}
and a normalisation factor 
\begin{align}
N(t)=\frac{1}{\sqrt{\left\|\bigg(\frac{t}{T} \mathcal{M} + (1-\frac{t}{T}) I \bigg)\ket{\psi_0}\right\|}}.
\end{align}
The normalisation factor $N(t)$ can be measured from the  expectation values of $\mathcal{M}^\dag+\mathcal{M}$ and $\mathcal{M}^\dag\mathcal{M}$ for $\ket{\psi_0}$. Given the definition of the state $\ket{\psi(t)}$ at time $t$, the corresponding derivative equation is
\begin{align}
\frac{d}{dt} \ket{\psi(t)}= \frac{\dot{N}(t) }{N(t)} \ket{\psi(t)}+N(t) G \ket{\psi(0)}.
\label{equ}
\end{align}
Such an equation is also described by the generalised time evolution equation with $\ket{v(t)}=\ket{\psi(t)}$, $A_1(t)=\frac{\dot{N}(t)}{N(t)} I$, $A_2(t)=N(t)G$, $\ket{v'_1(t)}=\ket{\psi(t)}$, $\ket{v'_2(t)}=\ket{\psi(0)}$, and $B(t)=1$.\\

\subsection{Solving linear equations}

We now discuss how to solve linear equations, defined by
$$\mathcal{M}\ket{v_{\mathcal{M}^{-1}}}=\ket{v_0},$$
 \red{with vector $\ket{v_0}$, invertible matrix $\mathcal M$, and solution $\ket{v_{\mathcal{M}^{-1}}}=\mathcal{M}^{-1}\ket{v_0}$}. 
We also introduce two algorithms for this problem where the first does not assume the structure of the matrix $\mathcal{M}$ and the second one assumes tensor product structure of $\mathcal{M}$. 
 
For the first type of algorithm, we convert the static problem into a dynamical problem. 
We
consider a path from  the initial state $\ket{v_0}$ to $\ket{v_{\mathcal{M}^{-1}}}$ as 
 $$E(t)\ket{v(t)}=\ket{v_0}$$ with 
 $$E(t)={t}/{T} \cdot \mathcal{M}+\big(1-{t}/{T}\big)I,$$ 
 $\ket{v(0)}=\ket{v_0}$, and $\ket{v(T)}=\ket{v_{\mathcal M^{-1}}}$.
That is, we can evolve the state $\ket{v_0}$ at time $t=0$ to the state $\ket{v_{\mathcal M^{-1}}}$ at time $t=T$.
The derivation equation of $\ket{v(t)}$ is given by 
 $$E(t)\frac{d}{d t}\ket{v(t)}=-G(t)\ket{v(t)}$$ with $G(t)=\big(\mathcal{M}-I\big)/T$. It is a special case of a generalised time evolution defined in Eq.~(\ref{Eq:general1}), with $B(t)=E(t)$, $A_j(t)=-G(t)$, and $\ket{v_j'(\vec{\theta}_j'(t))}=\ket{v(\vec{\theta}(t))}$. Therefore, we have the evolution equation as Eq.~\eqref{Eq:MVCEvol} with coefficients 
 $$\tilde M_{k,j} = \textrm{Re}\left(\frac{\partial \bra{v(\vec{\theta}(t))}}{\partial \theta_k}E^\dag(t)E(t)\frac{\partial \ket{v(\vec{\theta}(t))}}{\partial \theta_j}\right), \, \tilde V_{k} = -\textrm{Re}\left(\frac{\partial \bra{v(\vec{\theta}(t))}}{\partial \theta_k}E^\dag(t)G(t)\ket{v(\vec{\theta}(t))}\right).$$ 
 We can calculate $\mathcal{M}^{-1}\ket{v_0}$ by measuring the coefficients and evolving the parameters accordingly. It is important to note that $\tilde{V}_{k}$  can be efficiently computed with the quantum circuit shown in Fig. \ref{Fig:circuitPrac}. Therefore, linear equations can be solved with shallow quantum circuits which necessitate two or three controlled operations from an ancilla qubit.  

For the second type of algorithm, we assume that $\mathcal M=UDV$ and the solution is given by 
$$\ket{v_{\mathcal{M}^{-1}}}=V^\dag D^{-1}U^\dag\ket{v_0}.$$
With similar methods for realising matrix multiplication, we can also realise the matrix inversion operation with real and imaginary time evolutions.

\section{Resource estimation}
In this section, we discuss the complexity of the variational algorithms. We give the resource estimation for achieving a desired simulation accuracy.

\subsection{Resource estimation for simulating general processes}
We first discuss the resource required for implementing the variational algorithm for the generalised time evolution. Suppose we aim to simulate the evolution 
\begin{align}\label{general}
B(t) \frac{d}{dt}\ket{v(t)}=\sum_j A_j(t)\ket{v'_j(t)},
\end{align}
from time $t=0$ to $t=T$. We approximate $\ket{v(t)}$ by $\ket{v(\vec{\theta}(t))}$ and evolve the parameters according to 
\begin{align}
\sum_j \tilde{M}_{k,j} \dot{\theta}_j = \tilde{V}_k,
\end{align}
with a time step $\delta t$. Here the definition of the coefficients can be found in the main text. The error of the algorithm can be described by
\begin{equation}\label{Eq:}
\begin{aligned}
	  \varepsilon=D\left(\ket{v(T)},\ket{v(\vec{\theta}(T))}\right),\\
\end{aligned}
\end{equation}
where $D\left(\rho,\sigma\right)=\frac{1}{2}\tr\left[\left|\rho-\sigma\right|\right]$ is the trace distance between two states.
Denote $\mathcal{E}(t-\delta t,t)$ as the evolution from time $t-\delta t$ to $t$, then 
\begin{equation}\label{Eq:}
\begin{aligned}
	  \varepsilon&=D\left(\mathcal{E}(T-\delta t,T)(\ket{v(T-\delta t)}),\ket{v(\vec{\theta}(T))}\right),\\
	  &\le D\left(\mathcal{E}(T-\delta t,T)(\ket{v(T-\delta t)}),\mathcal{E}(T-\delta t,T)(\ket{v(\vec{\theta}(T-\delta t))})\right)+D\left(\mathcal{E}(T-\delta t,T)(\ket{v(\vec{\theta}(T-\delta t))}),\ket{v(\vec{\theta}(T))}\right),\\
	  &\le D\left(\ket{v(T-\delta t)},\ket{v(\vec{\theta}(T-\delta t))}\right)+D\left(\mathcal{E}(T-\delta t,T)(\ket{v(\vec{\theta}(T-\delta t))}),\ket{v(\vec{\theta}(T))}\right).\\
\end{aligned}
\end{equation}
Here the second line is due to the triangle inequality of the distance and the third line assumes that the evolution is dissipative in the sense that it can only decrease the distance. This is true for both real and imaginary time evolution and the open system simulation. Whether this is true for linear algebra problems depends on the problem. As this work mainly focuses on the simulation of open quantum systems, we leave a detailed discussion of the resource estimation for linear algebra problems in a future work. \\

Following a recursive procedure, we thus arrive at the final upper bound for the error of the simulation algorithm
\begin{equation}\label{Eq:}
\begin{aligned}
	  \varepsilon&\le \sum_{t=\delta t: \delta t: T}D\left(\mathcal{E}(t-\delta t,t)(\ket{v(\vec{\theta}(t-\delta t))}),\ket{v(\vec{\theta}(t))}\right),\\
\end{aligned}
\end{equation}
where we use $D\left(\ket{v(0)},\ket{v(\vec{\theta}(0))}\right)=0$ by assuming that the ansatz can perfectly represent the initial state at time $t=0$. For each term, we can also further divide it into two parts
\begin{equation}\label{Eq:}
\begin{aligned}
	  D\left(\mathcal{E}(t-\delta t,t)(\ket{v(\vec{\theta}(t-\delta t))}),\ket{v(\vec{\theta}(t))}\right)=\delta\varepsilon_I+\delta\varepsilon_A,\\
\end{aligned}
\end{equation}
with 
\begin{equation}\label{Eq:}
\begin{aligned}
	  \delta\varepsilon_I &=D\left(\ket{v(\vec{\theta}_0(t))},\ket{v(\vec{\theta}(t))}\right),\\
	  \delta\varepsilon_A &=D\left(\mathcal{E}(t-\delta t,t)(\ket{v(\vec{\theta}(t-\delta t))}),\ket{v(\vec{\theta}_0(t))}\right).\\
\end{aligned}
\end{equation}
Here $\ket{v(\vec{\theta}_0(t))}$ is the ideal state obtained with exact values of $\tilde M$ and $ \tilde V$. Therefore $\delta\varepsilon_I$ characterises the implementation error induced from imperfect $\tilde M$ and $ \tilde V$; $\delta\varepsilon_A$ characterises the algorithmic error due to finite time step and insufficient ansatz. Overall, the total error is upper bounded by
\begin{equation}\label{Eq:}
\begin{aligned}
	  \varepsilon&\le \varepsilon_I + \varepsilon_A = \sum_{t=\delta t: \delta t: T}(\delta\varepsilon_I+\delta\varepsilon_A),
\end{aligned}
\end{equation}
with $ \varepsilon_I=\sum_{t=\delta t: \delta t: T}\delta\varepsilon_I$ and $ \varepsilon_A=\sum_{t=\delta t: \delta t: T}\delta\varepsilon_A$.
In the following, we analyse these two terms separately.

\subsubsection{Implementation error}

The implementation error mainly comes from the imprecise estimation of $\tilde M$ and $ \tilde V$ owing to either physical error or shot noise. Denote the exact value of $\tilde M$ and $ \tilde V$ by $\tilde M_0$ and $ \tilde V_0$, respectively. Then we have the exact derivative $\dot{\vec{\theta}}_0=\tilde M_0^{-1}\tilde V_0$ and the practically measure derivative $\dot{\vec{\theta}}=\tilde M^{-1}\tilde V$. Denote $\tilde M=\tilde M_0+\delta\tilde M$, $ \tilde V= \tilde V_0+\delta  \tilde V$, and $\dot{\vec{\theta}}=\dot{\vec{\theta}}_0+\delta \dot{\vec{\theta}}$ with $\delta\tilde M$, $\delta  \tilde V$, and $\delta \dot{\vec{\theta}}$ representing the small noise perturbation. In Ref.~\cite{Li2017}, it is shown that the implementation error can be estimated by 
\begin{equation}\label{Eq:}
\begin{aligned}
	  \delta\varepsilon_I &=D\left(\ket{v(\vec{\theta}_0(t))},\ket{v(\vec{\theta}(t))}\right)=\sqrt{\delta \dot{\vec{\theta}}^T \mathcal{B} \delta \dot{\vec{\theta}} \delta t^2 +O(\delta t^3) },
\end{aligned}
\end{equation}
where 
\begin{align}
\mathcal{B}_{k,k'}=\frac{\partial \bra{v({\vec{\theta}(t)})}}{\partial \theta_k } \frac{\partial \ket{v({\vec{\theta}(t)})}}{\partial \theta_{k'} }-\frac{\partial \bra{v({\vec{\theta}(t)})}}{\partial \theta_k } \ket{v({\vec{\theta}(t)})}\bra{v({\vec{\theta}(t)})}\frac{\partial \ket{v({\vec{\theta}(t)})}}{\partial \theta_{k'} }.
\end{align}
Here we assumed that the evolution of Eq.~\eqref{general} is normalised, i.e.,  the state vector $\ket{v(\vec{\theta}(t))}$ is normalised.

The error of the parameters can be also represented as a function of $\delta\tilde M$ and $\delta  \tilde V$,
\begin{equation}\label{Eq:}
\begin{aligned}
\delta \dot{\vec{\theta}} &= \tilde M^{-1}\tilde V - \tilde M_0^{-1}\tilde V_0,\\
&=\big(\tilde{M}_0 (I+\tilde{M}_0^{-1} \delta \tilde{M}) \big)^{-1}	(\tilde{V}_0 +\delta \tilde{V})- \tilde M_0^{-1}\tilde V_0 \\
&=(I+\tilde{M}_0^{-1} \delta \tilde{M})^{-1}_0 \tilde{M}^{-1} (\tilde{V}_0+\delta \tilde{V})- \tilde M_0^{-1}\tilde V_0 \\
& \approx (I-\tilde{M}_0^{-1} \delta \tilde{M}) (\tilde{M}_0^{-1} \tilde{V}_0+\tilde{M}_0^{-1} \delta \tilde{V})- \tilde M_0^{-1}\tilde V_0 \\
&\approx \tilde{M}_0^{-1} \delta \tilde{V} - \tilde{M}_0^{-1} \delta \tilde{M} \dot{\vec{\theta}}_{0}.
\end{aligned}
\end{equation}
Here in the fourth line we consider a Taylor expansion with \textcolor{black}{$\|\tilde{M}_0 ^{-1} \delta \tilde{M}\| \leq \|\tilde{M}_0 ^{-1} \|  \|\delta \tilde{M}\| \ll 1$} and we omit higher order errors in the fifth line. 

Therefore we can upper bound the implementation error as
\begin{equation}\label{Eq:}
\begin{aligned}
	  \delta\varepsilon_I =\sqrt{\delta \dot{\vec{\theta}}^T \mathcal{B} \delta \dot{\vec{\theta}} \delta t^2 +O(\delta t^3) } &\lesssim \sqrt{\| \mathcal{B}\|} \|\delta \vec{\dot{\theta}}\|\delta t ,\\
	  &\lesssim \sqrt{\| \mathcal{B}\|} (\|\tilde{M}_0^{-1} \| \| \delta \tilde{V} \|+\|\tilde{M}_0^{-1} \|^2 \| \tilde{V}_0 \| \|\delta \tilde{M} \|)\delta t.
\end{aligned}
\end{equation}

Suppose the dominant error is the shot noise, we can have the relationship between the implementation error and the number of samples. Given the number of measurements $N_s$ for each term of $\tilde M$ and $ \tilde V$, we have
\begin{equation}
\begin{aligned}
 \|\delta \tilde{M} \| & \approx \frac{\sqrt{\sum_{j,k} \tilde{m}_{k,j}^2}}{\sqrt{N_S}},\quad 
 \|\delta \tilde{V} \| & \approx \frac{\sqrt{\sum_k \tilde{v}_k ^2}}{\sqrt{N_S}} 
\end{aligned}
\end{equation}
where 
\begin{equation}
\begin{aligned}
 \tilde{m}_{k,j}& = |\alpha (\vec{\theta}_0(t))|^2  \sum_{i,q} |g_{k,i}^* g_{j,q} | \|B(t) \|^2+ \bigg|\frac{\partial \alpha^* (\vec{\theta}_0(t) )}{\partial \theta_k}  \bigg| |\alpha(\vec{\theta}_0(t))|  \sum_i |g_{j,i}| \|B(t) \|\\
 &+\bigg|\frac{\partial \alpha^* (\vec{\theta}_0(t) )}{\partial \theta_j}  \bigg| |\alpha(\vec{\theta}_0(t))| \sum_i |g_{k,i}| \|B(t) \| +\bigg|\frac{\partial \alpha^* (\vec{\theta}_0(t) )}{\partial \theta_k}  \bigg| \bigg|\frac{\partial \alpha^* (\vec{\theta}_0(t) )}{\partial \theta_j}  \bigg| \|B(t) \|^2 \\
 \tilde{v}_k &= \sum_j \bigg[ \bigg|\frac{\partial \alpha^* (\vec{\theta}_0(t) )}{\partial \theta_k}  \bigg| | \alpha^* (\vec{\theta}_{0j}'(t) )|~ \|A_j(t) \|+ |\alpha^*(\vec{\theta}_0(t)) |~ | \alpha^* (\vec{\theta}_{0j}'(t) )| \sum_i|g_{k,i}|~ \| A_j(t) \| \bigg] \|B(t) \|,
\end{aligned}
\end{equation}
and $g_{k,i}$ comes from
\begin{equation}\label{Eq:decomposition}
\begin{aligned}
\frac{\partial \ket{\varphi(\vec{\theta}(t))}}{\partial \theta_k}&=\sum_i g_{k,i} R_{k,i} \ket{\bar{0}}.
\end{aligned}
\end{equation}

Therefore, $\| \delta\dot{ \vec{\theta}} \|$ can be upper bounded by 
\begin{align}
\|\delta \dot{\vec{\theta}} \|  \leq \frac{\Delta}{\sqrt{N_S}}
\end{align}
with 
\begin{align}
\Delta = \|\tilde{M}_0^{-1} \|^{-2} \|\tilde{V}_0 \|\sqrt{\sum_{j,k} \tilde{m}_{k,j}^2}+\|\tilde{M}_0^{-1} \| \sqrt{\sum_k \tilde{v}_k ^2}.
\end{align}
The implementation error at each step is thus upper bounded by
\begin{align}
\delta\varepsilon_I \leq \sqrt{\| \mathcal{B}\|} \frac{\Delta_{\mathrm{}} }
{\sqrt{N_S}}\delta t,
 \end{align}
and the total implementation error is
\begin{align}
\varepsilon_I \leq \sqrt{\| \mathcal{B}\|_{\mathrm{max}}} \frac{\Delta_{\mathrm{max}} }
{\sqrt{N_S}}T,
 \end{align}
 where $\| \mathcal{B}\|_{\mathrm{max}}$ and $\Delta_{\mathrm{max}}$ denote the maximal possible value at different times. Therefore, in order to have an implementation error $\varepsilon_I$,  we find that we can write
 $$N_S= \|\mathcal{B} \|_{\mathrm{max}} \Delta_{\mathrm{max}}^2 T^2 /\varepsilon_I^2.$$ 
 
\subsubsection{Algorithmic error}

The algorithmic error mainly comes from finite time step and imperfect ansatz. Following the definition of the time evolution, we have
\begin{equation}\label{Eq:}
\begin{aligned}
	  \delta\varepsilon_A &=D\left(\mathcal{E}(t-\delta t,t)(\ket{v(\vec{\theta}(t-\delta t))}),\ket{v(\vec{\theta}_0(t))}\right) = \sqrt{\Delta_2 \delta t^2 + \Delta_3 \delta t^3 +O(\delta t^4)}\lesssim\sqrt{\Delta_2^{(\mathrm{max})}} \delta t + 	\sqrt{\Delta_3^{(\mathrm{max})} \delta t} \delta t,\\
\end{aligned}
\end{equation}
where the first term comes from imperfection of ansatz and the second term comes from finite time step~\cite{Li2017}.
Here 
\begin{equation}
	\Delta_2=\braket{\delta v_1 (\vec{\theta})| \delta v_1 (\vec{\theta})}, \, \Delta_3=\braket{\delta v_1 (\vec{\theta})| \delta v_2 (\vec{\theta})}+\braket{\delta v_2 (\vec{\theta})| \delta v_1 (\vec{\theta})},
\end{equation}
with 
\begin{equation}
	\begin{aligned}
		 \ket{\delta v_1 (\vec{\theta}(t))} &= \ket{d v( t)} - \sum_j \dot{\theta}_j \frac{\partial}{\partial \theta_j} 	\ket{v( \vec{\theta}(t))} \\
 \ket{\delta v_2 (\vec{\theta}(t))} &=\frac{1}{2}\big(\frac{d}{dt} \ket{dv(t)} -\sum_{j j'} \dot{\theta}_j \dot{\theta}_{j'} \frac{\partial^2}{\partial \theta_j \partial \theta_{j'}} \ket{v(\vec{\theta}(t))}   \big).
	\end{aligned}	
\end{equation}
And $\Delta_2^{(\mathrm{max})}$ and $\Delta_3^{(\mathrm{max})}$ are the maximal possible values at time $t$, respectively.

Assuming that the ansatz can always represent the target state with $\Delta_2^{(\mathrm{max})}=0$, the total algorithmic error  is upper bounded by
\begin{equation}\label{Eq:}
\begin{aligned}
	 \varepsilon_A \lesssim\sqrt{\Delta_3^{(\mathrm{max})} \delta t} T.\\
\end{aligned}
\end{equation}
To suppress the effect of the second term to $\varepsilon_A$, we need to have $\delta t \approx \varepsilon_A ^{ 2}/ (\Delta_3^{(\mathrm{max} ) } T^2)$, hence the number of steps required is $N_A=T/\delta t \approx  \Delta_3^{(\mathrm{max} ) } T^3/ \varepsilon_A ^{ 2}$.

\subsubsection{Resource estimation}

Now, we combine the results and show the overall resource cost. At each step, the number of required individual quantum circuits $N_I$ is
\begin{align}
N_I = N_{B^\dag B}N_P^2 N_D^2 +2N_B N_D N_P +N_{B^\dag B} + N_B N_{A_j} N_{A_j}'(N_P N_D+1),    	
\end{align}
where the first term and the second terms correspond to the number of individual quantum circuits to populate each element of $\tilde{M}$ and $\tilde{V}$, respectively. Here, $N_P$ is the number of parameters, $N_D$ is the number of terms in Eq.~(\ref{Eq:decomposition}),  $N_{A_j}$, $N_{B}$ and $N_{B^\dag B}$ are the number of Pauli terms to decompose the $A_j (t)$, $B(t)$ and $B^\dag(t) B(t)$ matrices in Eq.~(\ref{general}), respectively, and $N_{A_j}'$ is the number of $A_j (t)$ matrices (the number of the terms in the right hand side of Eq. (\ref{general})). Thus the total number of measurements $N_{\mathrm{tot}}$ required to suppress the implementation error by shot noise to $\varepsilon_I$ and the effect of algorithmic error to $\varepsilon_A$ is 
\begin{equation}
\begin{aligned}
N_{\mathrm{tot}}&=N_I \times N_A \times N_S \\
&\approx\frac{\|\mathcal{B} \|_{\mathrm{max}} \Delta_{\mathrm{max}}^2 \Delta_3^{(\mathrm{max})}T^5}{\varepsilon_A^2\varepsilon_I^2} (N_{B^\dag B}N_P^2 N_D^2 +2N_B N_D N_P +N_{B^\dag B} + N_B N_{A_j} N_{A_j}'(N_P N_D+1)).
\end{aligned}
\end{equation}
By taking $\varepsilon_I=\varepsilon_A=\varepsilon/2$ with the total error $\varepsilon$, we have
\begin{equation}
\begin{aligned}
N_{\mathrm{tot}}&\approx N_I \times N_A \times N_S \\
&=\frac{16\|\mathcal{B} \|_{\mathrm{max}} \Delta_{\mathrm{max}}^2 \Delta_3^{(\mathrm{max})}T^5}{\varepsilon^4} (N_{B^\dag B}N_P^2 N_D^2 +2N_B N_D N_P +N_{B^\dag B} + N_B N_{A_j} N_{A_j}'(N_P N_D+1)).
\end{aligned}
\end{equation}
Note that this resource estimation is only a pessimistic asymptotic estimation, while the realistic resource required can be much less for a given problem further with the optimisation of measurement schemes

\subsection{Resource estimation for matrix multiplication via singular value decomposition method}
In this section, we show the resource required for realising matrix multiplication via the singular value decomposition method, which is used in the variational algorithm for simulating open quantum systems. We leave the detailed resource analysis for the other variational algorithms of linear algebra problems in a future work.

Suppose the matrix $\mathcal{M}$ can be decomposed as $\mathcal{M}=U D V$, where $U$ and $V$ are unitary matrices and $D$ is a diagonal matrix with non-negative entries. Suppose that $U$, $V$ and $D$ can be decomposed as $U=\mathrm{exp}(-iH^U T^U)$ and $V=\mathrm{exp}(-iH^V T^V)$ and $D \approx \mathrm{exp}(-H^D T^D)$, and these operations can be realised by variational real and imaginary time simulation algorithms. As we only approximate  $D=\sum_j a_j \ket{j} \bra{j}$ by $D \approx \mathrm{exp}(-H^D T^D)$ with 
$-H^{D}T = \sum_{a_j\neq 0} \mathrm{log} (a_j) \ket{j} \bra{j} -\alpha  \sum_{a_j=0}  \ket{j}\bra{j}$, we first analyse the error introduced from this approximation.

\subsubsection{Accuracy of approximating $D$}
Denote $D_\alpha=D+\Delta_\alpha$ with $D=\sum_{a_j \neq } a_j \ket{j}\bra{j}$ and $\Delta_\alpha=\sum_{a_j=0}e^{-\alpha}\ket{j}\bra{j}$. We use $D_\alpha$ to approximate $D$, for a given vector $\ket{v}$, the error of the matrix can propagate to the state as
\begin{equation}
	\begin{aligned}
		\varepsilon_D=&\left\|\frac{D_\alpha\ket{v}\bra{v}D_\alpha}{\bra{v}D_\alpha^2\ket{v}}- \frac{D\ket{v}\bra{v}D}{\bra{v}D^2\ket{v}}\right\|_1 \\
		=& \frac{1}{\bra{v}D^2\ket{v}}\left\|\frac{\bra{v}D^2\ket{v}}{\bra{v}D_\alpha^2\ket{v}}D_\alpha\ket{v}\bra{v}D_\alpha- {D\ket{v}\bra{v}D}\right\|_1,\\
		=&\frac{1}{\bra{v}D^2\ket{v}}\left\|\frac{\bra{v}D^2\ket{v}}{\bra{v}D_\alpha^2\ket{v}}(D+\Delta_\alpha)\ket{v}\bra{v}(D+\Delta_\alpha)- {D\ket{v}\bra{v}D}\right\|_1,\\
		=&\frac{1}{\bra{v}D^2\ket{v}}\left\|\left(\frac{\bra{v}D^2\ket{v}}{\bra{v}D_\alpha^2\ket{v}}-1\right){D\ket{v}\bra{v}D}+\frac{\bra{v}D^2\ket{v}}{\bra{v}D_\alpha^2\ket{v}}\left( (D+\Delta_\alpha)\ket{v}\bra{v}(D+\Delta_\alpha)-{D\ket{v}\bra{v}D}\right)\right\|_1,\\
		\le& \left|\frac{\bra{v}D^2\ket{v}}{\bra{v}D_\alpha^2\ket{v}}-1\right| + \frac{1}{\bra{v}D^2\ket{v}}\left\|(D+\Delta_\alpha)\ket{v}\bra{v}(D+\Delta_\alpha)-{D\ket{v}\bra{v}D}\right\|_1.\\
	\end{aligned}
\end{equation}
Here $\|\rho\|_1=\tr|\rho|$ is the trace norm of $\rho$. The first term can be bounded as
\begin{equation}
	\begin{aligned}
		\left|\frac{\bra{v}D^2\ket{v}}{\bra{v}D_\alpha^2\ket{v}}-1\right| &= \left|\frac{\bra{v}D_\alpha^2\ket{v}-\bra{v}D^2\ket{v}}{\bra{v}D_\alpha^2\ket{v}}\right|=\left|\frac{\bra{v}\Delta_\alpha^2\ket{v}}{\bra{v}D_\alpha^2\ket{v}}\right|\le \frac{e^{-2\alpha}}{C},\\
	\end{aligned}
\end{equation}
where $C=\bra{v}D^2\ket{v}$ determines the norm after multiplying the matrix to the vector, which can be assumed to be lower bounded by a constant. As the value of $C$ can be experimentally measured, the cases with an exponential small $C$ can be experimentally identified. In the case with an exponentially small $C$, it just indicate that the output vector is a zero vector. 

For the second term, we can bound it as
\begin{equation}
	\begin{aligned}
		&\frac{1}{\bra{v}D^2\ket{v}}\left\|(D+\Delta_\alpha)\ket{v}\bra{v}(D+\Delta_\alpha)-{D\ket{v}\bra{v}D}\right\|_1 \\
		\le & \frac{1}{C}\left\|\Delta_\alpha \ket{v}\bra{v}D+D \ket{v}\bra{v} \Delta_\alpha+\Delta_{\alpha}\ket{v}\bra{v}\Delta_\alpha\right\|_1 \\
		\le & \frac{1}{C}\left\|\Delta_{\alpha} \ket{v}\bra{v}D\right\|_1+\left\|D\ket{v}\bra{v}\Delta_\alpha\right\|_1+\left\|\Delta_\alpha\ket{v}\bra{v}\Delta_\alpha\right\|_1 \\
		=&\frac{e^{-2\alpha}}{C},
	\end{aligned}
\end{equation}
{where we used $\left\|\Delta_{\alpha} \ket{v}\bra{v}D\right\|_1=\left \|D\ket{v}\bra{v}\Delta_\alpha \right \|_1=0$.}
Therefore, the total error induced from the approximation of $D_\alpha$ is
\begin{equation}
	\varepsilon_D = \frac{2e^{-2\alpha}}{C},
\end{equation}
and we can choose $\alpha=\ln(2/C\varepsilon_D)/2$. Therefore, when $C\ge 1/\textrm{Poly}(n)$ with $n$ denoting the number of qubits, we have 
\begin{equation}
	\alpha = O\left(\log\left(\frac{1}{\varepsilon_D}\right)+\log\textrm{Poly}(n)\right).
\end{equation}
We note that even when $C$ is exponentially small with respect to $n$ as $C=1/c^n$ with constant  $c$, we can still choose $\alpha$ to be a constant. Nevertheless,  we should always measure the value of $C$ at the beginning of the algorithm and directly output a zero vector when $C$ is too small.

\subsubsection{Resource analysis}
According to our resource analysis for the generalised time evolution together with the error for approximating $D$, the implementation error is 
\begin{align}
\varepsilon_I &\lesssim \sqrt{\|\mathcal{B} \|_{\mathrm{max}}} \frac{\Delta_{\mathrm{max}}T_{SVD}}{\sqrt{N_S}},
\end{align}
and the algorithmic error is 
\begin{align}
\varepsilon_A &\lesssim \sqrt{\Delta_3^{(\mathrm{max})} \delta t} T_{SVD}+\varepsilon_D, 
\end{align}
where $ T_{SVD}=T^U+T^V+T^D$. 
Therefore, the total number of measurements $N_{\mathrm{tot}}^{SVD}$ required to suppress the effect of shot noise to $\varepsilon_I$ and the effect of algorithmic error to $\varepsilon_A$ is
\begin{equation}
\begin{aligned}
N_{\mathrm{tot}}^{SVD} \approx \frac{\|\mathcal{B} \|_{\mathrm{max}} \Delta_{\mathrm{max}}^2 \Delta_3^{(\mathrm{max})}T_{SVD}^{~5}}{(\varepsilon_A-\varepsilon_D)^2\varepsilon_I^2} (N_P^2 N_D^2 + N_P N_H N_D),
\end{aligned}
\end{equation}
where $N_H$ is the largest number of terms when $H^{U}$, $H^{V}$, and $H^{D}$ is decomposed as a linear combination of Pauli operators.

\subsection{Resource estimation for stochastic Sch\"odinger equation}

The resource estimation for the variational algorithm of open quantum systems is  more involved. The simulation error consists of the following parts
\begin{enumerate}
	\item The algorithmic error from approximating the Lindblad master equation with the stochastic Schr\"odinger equation with finite number of trajectory samples.
	\item For each trajectory, we approximate the continuous evolution with the generalised time evolution and the jump process with the matrix multiplication algorithm.
	\item For each jump, we need to estimate the jump probability and determining the jump time, whose error can also cause implementation errors. 
\end{enumerate}

For the first type of error, the error can be upper bounded to a small value with a sufficiently large number of sample proportional to a polynomial function of the inverse of the accuracy.  
For the second type of error, we can bound the error with the analysis for the generalised time evolution. For the jump process, we can use the analysis for matrix multiplication for each jump and we show in the following that the number of jumps is proportional to the evolution time. For the third type error, we can also bound it with a similar analysis for the generalised time evolution. 
In the following, we show how to estimate the number of jumps and we leave the detailed analysis of resource estimation in a future work.

\subsubsection{Number of jumps}
One trajectory of the stochastic Schr\"odinger equation is composed of the continuous evolution and jumps processes. The resource cost of the continuous evolution is similar to the one for real time evolution discussed in Ref.~\cite{Li2017}, which is shown to be polynomial to the evolution time and system size. For the jump processes, as each jump is simulated with real and imaginary time evolution, the resource cost of each jump is therefore also polynomial to the evolution time and system size.

Then we discuss how many jump process occur on average in the simulation of the Stochastic Schr\"odinger equation. The averaged number of jump events during time from $t$ to $t+dt$ is 
\begin{align}
\bra{\psi_c(t)}\sum_k L_k^\dag L_k \ket{\psi_c(t)} dt.
\end{align}
Therefore, the average number of jump events from $t=0$ to $t=T$
\begin{equation}
	\begin{aligned}
		N_{\mathrm{jump}} &= \int_0 ^T \bra{\psi_c(t)}\sum_k L_k^\dag L_k \ket{\psi_c(t)} dt,\\
		&\le   T \left\|\sum_k L_k^\dag L_k\right\|_\infty,\\
		&\le    T \sum_k\left\| L_k^\dag L_k\right\|_\infty,
	\end{aligned}
\end{equation}
\textcolor{black}{where $\|L\|_\infty$ is the operator norm, which is the largest singular value of $L$.} For physical systems, we generally have $\|L_k^\dag L_k\|_\infty$ and the number of Lindblad terms equal $O(\textrm{Poly}(n))$, where $n$ is the system size and $\textrm{Poly}(n)$ is a polynomial function of $n$. Therefore, the averaged number of jumps is 
\begin{equation}
	N_{\mathrm{jump}}  = O(T\cdot\textrm{Poly}(n)).
\end{equation}

The number of jumps is much fewer when considering the case where each Lindblad operator only locally acts on a constant subsystem. That is, we assume that $L^\dag_k L_k $ has orthogonal support to each other and $\|L^\dag_k L_k\|=O(1)$. In this case, we have $\left\|\sum_k L_k^\dag L_k\right\|_\infty=O(1)$ and hence 
\begin{equation}
	N_{\mathrm{jump}}  = O(T).
\end{equation}

To simulate the stochastic Schr\"odinger equation, we also need to sample different random trajectories. 
When we measure an observable $O$ and hope to suppress the sampling error to $\epsilon=1/\sqrt{M}$, we need $M$ random samples. Therefore, the overall cost should also be multiplied by $M$. However, it is worth noting that every trajectory is exactly parallel \textcolor{black}{so the overhead $M$ can be also reduced by a constant factor.}

\end{document}